\documentclass{amsart}
\usepackage{amsmath}
\usepackage{amsthm}
\begin{document}

\title{Spline-interpolation solution of 3D Dirichlet problem for one class of solids}

\author{P.N.Ivanshin}
\address{Physics Institute, Kazan Federal
University, Kazan, 420008, Russia}
\email{pivanshin@gmail.com}
\author{E.A.Shirokova}
\address{Mathematics \& Mechanics Institute, Kazan Federal
University, Kazan, 420008, Russia}
\email{Elena.Shirokova@ksu.ru}






\maketitle

\begin{abstract}
We present the spline-interpolation approximate solution of the
Dirichlet problem for the Laplace equation in the bodies of
revolution, cones and cylinders. Our method is based on reduction of
the 3D problem to the sequence of 2D Dirichlet problems. The main
advantage of the spline-interpolation solution of the 3D Dirichlet
problem is its continuity in the whole domain up to the boundary
even for the case of the linear spline.
\end{abstract}

\section{Introduction}

We present the new --- spline-interpolation --- approximate solution
of the Dirichlet problem for the Laplace equation  which has wide
applications to electric and seismic prospecting \cite{Back, CMM,
DL, Keller, Nei, Wen}. This method is applicable for the cylinders,
cones and the bodies of revolution. There exist many methods of
approximate solution of this problem. The classic approximate method
is the finite-element one. The spline-interpolation method is the
analog of finite-element method where the discrete boundary nodes
are the closed boundary curves and the cells are the layers between
two parallel planes. Our method is based on reduction of the 3D
problem to the sequence of 2D Dirichlet problems. Note that one can
apply the functions of complex variable in the case of 2D Dirichlet
problem. There also exists another spline technique of solution, but
the one presented here is substantially different from the one
given, for example, in the book \cite{Iso}. The main advantage of
the spline-interpolation solution of the 3D Dirichlet problem is its
continuity in the whole domain up to the boundary even for the case
of the linear spline. The piece-wise analytic form of the
spline-interpolation solution is also convenient for the
applications.

\section{Formulation of the problem}

Let $M$ be a body of revolution with the smooth generatrix
$x=\Phi(h)$ in $XYH$ space.  Let us denote by $S$ the boundary of
$M$.

The classic internal Dirichlet problem for the Laplace equation
\begin{eqnarray}
\Delta_3 u=0,\label{eq:1}
\end{eqnarray}
is the following: given the function $U_0$ on $S$ it should be found
the function $u$ such that it satisfies equation (\ref{eq:1})
everywhere in $M$ and takes the given boundary values $U_0$, that is
\begin{eqnarray}
u|_{S}=U_0.\nonumber
\end{eqnarray}
Here
\begin{eqnarray}
\Delta_3=\partial^2_x+\partial^2_y+\partial^2_h.\nonumber
\end{eqnarray}

We call by the spline-interpolation solution of the problem the
function which satisfies equation (\ref{eq:1}) almost everywhere in
$M$ and takes the given continuous boundary values at a finite
number of the curves being the sections of $S$ by the planes
$h=h_j$. So we do not solve the initial boundary value problem but
reduce it to the problem with the data given at some finite set of
sections of $S$. For the case when the domain $M$ has the upper and
the lower ends $D_A$ and $D_B$ on the planes $h=A$ and $h=B$ we
assume that the given boundary values at these ends are polyharmonic
functions. The spline-interpolation solution takes these given
values at the ends. Note that the requirement of poly-harmonicity is
not too restrictive because every continuous function given on the
plane can be approximated with any accuracy and any smoothness by a
polyharmonic function.


It is well-known \cite{Vlad} that the external problem  can be
reduced to the internal one  by Kelvin transform \cite{Kel}. Let
$\mathbb{R}^3\setminus M$ be the unbounded domain in which we must
solve equation (\ref{eq:1}). We assume that the unit ball $B_1(0)$
is a subset of $M$. Then the inversion of $\mathbb{R}^3\setminus M$
with respect to the unit sphere belongs to $B_1(0)$.  It is
well-known, that $u(x, y, h)$ is the solution of equation
(\ref{eq:1}) inside the  ball $B_1(0)$ if and only if the function
$v(x,y,h)=$ $u(x/r^2, y/r^2, h/r^2)/r$, $r=\sqrt{x^2+y^2+h^2}$ is
the solution of (\ref{eq:1}) outside the ball $B_1(0)$ which
vanishes at infinity ($O(\frac{1}{r})$). So we present and
demonstrate the spline-interpolation solution of the internal
problem for the body of revolution and only then discuss the case of
the external problem and the problem for cylinders or cones.

\section{Analysis}

We construct the spline-interpolation solution for every fragment of
$M$ which is the layer between two planes $h=h_j$, $h=h_{j+1}$. The
form of spline-interpolation solution is polynomial with respect to
$h$. If we put the solution in the form
\begin{eqnarray}
u(x,y,h)=\sum\limits_{k=0}^{p} h^k u_k(x, y),\nonumber
\end{eqnarray}
in equation (\ref{eq:1}) and equate the coefficients with the same
powers of $h$ we get the following relations:
\begin{eqnarray}
(k+2) (k+1)u_{k+2}+\Delta_2 u_{k}=0, \, k=0, \ldots, p-2,\nonumber
\end{eqnarray}
and
\begin{eqnarray}
\Delta_2 u_k=0, \, k=p, p-1,\nonumber
\end{eqnarray}
where
\begin{eqnarray}
\Delta_2=\partial^2_x+\partial^2_y.\nonumber
\end{eqnarray}

Now the solution $u(x,y,h)$ satisfies equation (\ref{eq:1}) in the
layer and $u(x,y,h)$ is a $([\frac{p}{2}]+1)$-harmonic function for
any fixed $h$. The main problem now is to satisfy the boundary
condition and to glue continuously the solutions constructed for the
adjacent layers. Such spline-interpolation solution is named here
the continuous solution. The derivative $u_h(x,y,h)$ of the
continuous solution is bounded but discontinuous at the points of
$M$ on the cuts $h=h_j$. We also show how to minimize these
discontinuity.

\section{The continuous solution. The case of a domain without ends}

Assume that the body of revolution $M$ lies between the planes $h=A$
and $h=B$, the vertices $P_A$ and $P_B$ being the common points of
the domain and the corresponding planes. Assume that the boundary
values are given at $m$ circles $C_j$, $j=1,...,m,$ which are the
sections of $S$ by the planes $h=h_j$, $j=1,...,m,$ in the form
$f_j: C_j \to \bf{R}$ and also at the points $P_A$ and $P_B$.

We begin the construction of the spline-interpolation solution
within the layer which contains one of the vertices, namely $P_A$,
so it is the layer $h \in [A,h_1]$. Here we put
\begin{eqnarray}
u^1(x,y,h)=u^1_0(x,y)+(h-h_1) u^1_1,\nonumber
\end{eqnarray}
where $u^1_0(x,y)$ is a harmonic function in the domain which is the
section of $M$ by the plane $h=h_1$ and $u^1_1$ is a constant.
Evidently this spline satisfies equation (\ref{eq:1}). The function
$u^1_0(x,y)$ can be restored via the boundary values $f_1$. We find
the constant $u^1_1$ using the relation
\begin{eqnarray}
u^1_1=\frac{u(P_A)-u^1_0(x_A,y_A)}{(A-h_1)},\nonumber
\end{eqnarray}
where $u(P_A)$ is the given value at the vertex $P_A$, the
parameters $x_A$ and $y_A$ are the abscissa and the ordinate of the
point $P_A$, respectively.

In the next layer $h \in[h_1,h_2]$ the spline has the form
\begin{eqnarray}
u^2(x,y,h)=u^2_0(x,y)+(h-h_1) u^2_1(x,y),\nonumber
\end{eqnarray}
where both of the unknown functions $u^2_0(x, y)$ and $u^2_1(x, y)$
are harmonic in the union of the projections of the sections of $M$
by the planes  $h=h_1$, $h=h_2$ onto the plane $XOY$. Evidently this
spline satisfies equation (\ref{eq:1}).

We put $u^2_0(x, y) = u^1_0(x, y)$. So the boundary conditions at
the curve $C_1$ are met. Now we find the boundary values of the
function $u^2_1$ which are equal  to
\begin{eqnarray}
\frac{f_2-u^2_0(C_2)}{h_2-h_1},\nonumber
\end{eqnarray}
and restore the harmonic in the section of $M$ by the plane $h=h_2$
function $u^2_1(x,y)$ via its boundary values.

In the layer $h \in[h_2,h_3]$ we put
\begin{eqnarray}
u^3(x,y,h)=u^3_0(x,y)+(h-h_2) u^3_1(x,y),\nonumber
\end{eqnarray}
where  the harmonic function $u^3_0(x,y)$ can be restored by setting
$u^3_0(x,y)=u^2_0(x,y)+(h_2-h_1)u^2_1(x,y)$, and the harmonic
function $u^3_1(x,y)$ can be restored via the boundary values
\begin{eqnarray}
\frac{f_3-u^3_0(C_3)}{h_3-h_2}.\nonumber
\end{eqnarray}

We move from one layer to the next one successively and restore the
functions $u^k_0(x,y)$, $u^k_1(x,y)$, $k=1,...,m$ in the whole
domain $M$. The construction for the last layer is similar to that
for the first one.

Note that the harmonic on one cut of $M$ by $h=h_k$ function must be
continuous and harmonic at the neighbor cuts. It holds true if  the
boundary data on the curves $C_j$, $j=1,...,m$, is given in the form
of  trigonometric polynomials.

The constructed spline-interpolation solution is harmonic at the
points of the common plane of two adjacent layers. The corresponding
function is restored uniquely via its boundary values, so this
solution is continuous over the axis $OH$. It is also continuous in
any plane parallel to $XOY$ up to the boundary of $M$. So the
constructed linear with respect to $h$ spline-interpolation solution
is continuous in $M$ up to the boundary.

The derivatives of the constructed solution with respect to $x$ and
to $y$ are continuous in $M$ at every level $h=const$, but the first
derivative with respect to $h$ is discontinuous at the points of $M$
at the levels $h=h_j$, $j=1,...,m$. We show that this discontinuity
tends to zero when $m \to \infty$ and
$\max\limits_{j}|h_{j+1}-h_j|\to 0$ if the boundary data of the
corresponding classic Dirichlet problem is a smooth function and the
boundary surface of $M$ is smooth.

Suppose that the boundary data of the classic Dirichlet problem is
the function $f(\theta, h)$, $\theta\in [0,2\pi]$, $h\in [A,B]$,
which has the continuous derivative $f_h(\theta,h)$, which belongs
to H\"older class $H(\alpha)$ with respect to $h$ uniformly for
$\theta\in [0,2\pi]$. So there exist the constants $L>0$ and
$\alpha\in (0,1]$ such that $|f'_h(\theta,h')-f'_h(\theta,h'')|\leq
L |h'-h''|^\alpha$ for any $\theta\in [0,2\pi]$. Let the generatrix
$x=\Phi(h)$ of the boundary surface of $M$ be smooth: $\Phi'(h)$
belongs to $H(\alpha)$. We assume that the distance between any two
adjacent planes $h=h_j$ equals $t=(B-A)/(m+1)$. Consider two layers
in $M$ with the common plane $h=0$ and the splines constructed for
these layers:
$$
u^j(x,y,h)=u_0(x,y)+u^j_1(x,y) h, \   h\in [-t,0],
$$
and
$$
u^{j+1}(x,y,h)=u_0(x,y)+u^{j+1}_1(x,y) h, \    h\in [0,t].
$$

We have
$$
u^j(\Phi(-t)\cos\theta,\Phi(-t)\sin\theta,-t)=f(\theta,-t),
$$
$$
u^j(\Phi(0)\cos\theta,\Phi(0)\sin\theta,0)=
u^{j+1}(\Phi(0)\cos\theta,\Phi(0)\sin\theta,0)=f(\theta,0),
$$
$$
u^{j+1}(\Phi(t)\cos\theta,\Phi(t)\sin\theta,t)=f(\theta,t),
$$
so
$$
u_0(\Phi(0)\cos\theta,\Phi(0)\sin\theta,0)=f(\theta,0),
$$
$$
-t u^j_1(\Phi(-t)\cos\theta,\Phi(-t)\sin\theta)=
[f(\theta,-t)-f(\theta,0)]+
$$
$$
+[u_0(\Phi(0)\cos\theta,\Phi(0)\sin\theta,0)-
u_0(\Phi(-t)\cos\theta,\Phi(-t)\sin\theta)],
$$
$$
t u^{j+1}_1(\Phi(t)\cos\theta,\Phi(t)\sin\theta)=
[f(\theta,t)-f(\theta,0)]+
$$
$$
+[u_0(\Phi(0)\cos\theta,\Phi(0)\sin\theta,0)-
u_0(\Phi(t)\cos\theta,\Phi(t)\sin\theta)].
$$

We apply Lagrange formula and have
$$
u^j_1(\Phi(-t)\cos\theta,\Phi(-t)\sin\theta)=
f'_t(\theta,\eta(\theta) t)-[u'_{0 x}(\Phi(\mu(\theta) t)\cos\theta,
\Phi(\mu(\theta) t)\sin\theta) \cos\theta+
$$
$$
+u'_{0 y}(\Phi(\mu(\theta) t)\cos\theta, \Phi(\mu(\theta)
t)\sin\theta) \sin\theta] \Phi'(\mu(\theta) t),\
\eta(\theta),\mu(\theta)\in(-1,0),
$$
$$
u^{j+1}_1(\Phi(t)\cos\theta,\Phi(t)\sin\theta)=
f'_t(\theta,\xi(\theta) t)-[u'_{0 x}(\Phi(\mu(\theta) t)\cos\theta,
\Phi(\mu(\theta) t)\sin\theta) \cos\theta+
$$
$$
+u'_{0 y}(\Phi(\mu(\theta) t)\cos\theta, \Phi(\nu(\theta)
t)\sin\theta) \sin\theta] \Phi'(\nu(\theta) t),\
\xi(\theta),\nu(\theta)\in(0,1).
$$
Note that the functions $u'_{0 x}(x,y)$ and $u'_{0 y}$ are
differentiable in the union of the disks with the radii $\Phi(-t)$,
$\Phi(0)$ and $\Phi(t)$.

Now
$$
u^j_1(\Phi(0)\cos\theta,\Phi(0)\sin\theta)=u^j_1(\Phi(-t)\cos\theta,\Phi(-t)\sin\theta)+
$$
$$
+[u^j_1(\Phi(0)\cos\theta,\Phi(0)\sin\theta)-u^j_1(\Phi(-t)\cos\theta,\Phi(-t)\sin\theta)],
$$
$$
u^{j+1}_1(\Phi(0)\cos\theta,\Phi(0)\sin\theta)=u^{j+1}_1(\Phi(t)\cos\theta,\Phi(t)\sin\theta)+
$$
$$
+[u^{j+1}_1(\Phi(0)\cos\theta,\Phi(0)\sin\theta)-u^{j+1}_1(\Phi(t)\cos\theta,\Phi(t)\sin\theta)].
$$

Finally we have according to supposition
$$
|u^{j+1}_1(\Phi(0)\cos\theta,\Phi(0)\sin\theta)-u^j_1(\Phi(0)\cos\theta,\Phi(0)\sin\theta)|\leq
K t^{\alpha}.
$$

The function $u^{j+1}_1(x,y)-u^{j}_1(x,y)$ is harmonic in the disk
$x^2+y^2\leq \Phi(0)$, so $|u^{j+1}_1(x,y)-u^{j}_1(x,y)|$ $\leq K
t^{\alpha}$ everywhere in this disk, and we can see that the
discontinuity of the function $u_h(x,y,h)$ vanishes when the
distance $t$ between any two adjacent planes $h=h_j$ tends to zero.

We have the representation
$$
u^{j+1}_1(\Phi(0)\cos\theta,\Phi(0)\sin\theta)-u^j_1(\Phi(0)\cos\theta,\Phi(0)\sin\theta)=
t \phi(\theta,t)
$$
where $\phi(\theta,t)$ is bounded when $t$ tends to zero if the
functions $\Phi''(h)$ and $f_{hh}(\theta,h)$ are continuous.

Note that we can construct the solution beginning from the lowest
point as well as from the highest point. If $u_i(x,y,h)$ and
$u_f(x,y,h)$ are such solutions then $\alpha u_i(x,y,h)+(1-\alpha)
u_f(x,y,h)$ is also the solution for any $\alpha \in(0,1)$.

\section{The continuous solution. The case of a domain with the ends}

Here the domain $M$ is situated between the planes $h=A$ and $h=B$,
the end $D_A$ is located on the plane $h=A$ and the end $D_B$ is
located on the plane $h=B$, the curves $C_j$ are located on the
planes $h=h_j$, $j=1,...,m,$. The boundary data on the ends has the
form $U_A(x,y)$, $(x,y) \in E_A$ and $U_B(x,y)$,$(x,y) \in E_B$. The
boundary values at the points of the curves $C_j$ are given by the
functions  $f(\theta,h_j)$.

The form of the solution depends on the order of harmonicity of the
functions $U_A(x,y)$ and $U_B(x,y)$. Assume that the highest order
of harmonicity of these functions is equal to $(n+1)$.

We construct the spline for the layer $h \in [A,h_1]$ in the form
\begin{eqnarray}
u^1(x,y,h)=u^1_0(x,y)+(h-A) u^1_1(x,y) + \sum_{k=1}^n u_{2
k}^1(x,y)\nonumber (h-A)^{2 k},
\end{eqnarray}
where $u^1_0(x,y)=U_A(x,y)$. The spline satisfies equation
(\ref{eq:1}) if and only if
\begin{eqnarray}
u_{2 k}^1(x,y)=\frac{(-1)^k}{(2 k)!} \Delta_2^{k}u^1_0(x,y),
k=1,...,n,\nonumber
\end{eqnarray}
where $\Delta_2^{k}$ is $k$ times applied Laplace operator
$\Delta_2=\partial^2_x+\partial^2_y$. Hence we find all coefficients
with the even powers of $h$ if we satisfy equation (\ref{eq:1}) in
the layer $h \in [A,h_1]$. The only odd coefficient $u^1_1(x,y)$
must satisfy the equation
\begin{eqnarray}
\Delta_2 u^1_1(x,y)=0.\nonumber
\end{eqnarray}

Therefore we have to restore the function $u^1_1(x,y)$ which is
harmonic in the union of the projections of the sections of $M$ by
the planes $h=A$, $h=h_1$ onto $XOY$. We do it using the boundary
condition
\begin{eqnarray}
u^1_1(x,y)|_{C_1}=\frac{f_1-u^1_0(C_1)-\sum_{k=1}^n u_{2
k}^1(C_1)}{h_1-A}.\nonumber
\end{eqnarray}

For the next layer $h \in [h_1,h_2]$ we take the spline in the form
\begin{eqnarray}
u^2(x,y,h)=u^2_0(x,y)+(h-h_1) u^2_1(x,y) + \sum_{k=1}^n u_{2
k}^2(x,y) (h-h_1)^{2 k},\nonumber
\end{eqnarray}
where
\begin{eqnarray}
u^2_0(x,y)=u^1_0(x,y)+(h_1-A) u^1_1(x,y) + \sum_{k=1}^n u_{2
k}^1(x,y) (h_1-A)^{2 k},\nonumber
\end{eqnarray}
and
\begin{eqnarray}
u_{2 k}^2(x,y)=\frac{(-1)^k}{(2 k)!} \Delta_2^{k}u^2_0(x,y),
k=1,...,n.\nonumber
\end{eqnarray}

The harmonic function $u^2_1(x,y)$ can be restored via the boundary
data at the curve $C_2$.

All the splines for the next layers are constructed similarly. Only
the last spline where we must satisfy both the given data on the
section and on the end $D_B$ is different.

We take the last spline for the layer $h \in [h_m,B]$ in the form
\begin{eqnarray}
u^m(x,y,h)=u^m_0(x,y)+(h-h_m) u^m_1(x,y)+ \nonumber \\
+ \sum_{k=1}^n u_{2 k}^m(x,y) (h-h_m)^{2 k} + \sum_{k=1}^n u_{2
k+1}^m(x,y) (h-h_m)^{2 k+1},\nonumber
\end{eqnarray}
where
\begin{eqnarray}
u^m_0(x,y)=u^{m-1}_0(x,y)+(h_m-h_{m-1}) u^{m-1}_1(x,y) +\nonumber\\
+ \sum_{k=1}^n u_{2 k}^{m-1}(x,y) (h_m-h_{m-1})^{2 k},\nonumber
\end{eqnarray}
\begin{eqnarray}
u_{2 k}^m(x,y)=\frac{(-1)^k}{(2 k)!}
\Delta_2^{k}u^m_0(x,y).\nonumber
\end{eqnarray}

So we glue continuously the last spline with the previous one at the
points of the common section and provide equality of the
coefficients with the even powers of $h$ in (\ref{eq:1}). Now we
must provide equality of the coefficients with the odd powers of $h$
in (\ref{eq:1}), that is why we put
\begin{eqnarray}
u_{2k+1}^m(x,y)=\frac{(-1)^{k}}{(2k+1)!} \Delta_2^{k} u^m_1(x,y),\
k=1,...,n.\nonumber
\end{eqnarray}

Now equation (\ref{eq:1}) is satisfied in the layer $[h_m,B]$ and we
must satisfy the boundary data at the points of  $D_B$. We write
this boundary condition in the form
\begin{eqnarray}
\sum\limits_{k=0}^{n}\frac{(-1)^{k}}{(2k+1)!} \Delta_2^{k}
u^m_1(x,y) (B-h_m)^{2 k+1} =\nonumber\\
=U_B(x,y)-\sum\limits_{k=0}^{n}\frac{(-1)^{k}}{(2k)!} \Delta_2^{k}
u^m_0(x,y) (B-h_m)^{2 k+1},\nonumber
\end{eqnarray}
where the right side of the relation is known $(n+1)$-harmonic
function, we denote it by $F_B(x,y)$. Our aim is to restore the
$(n+1)$-harmonic function $u^m_1(x,y)$ with the help of the identity
\begin{eqnarray}
\sum\limits_{k=0}^{n}\frac{(-1)^{k}}{(2k+1)!} \Delta_2^{k}
u^m_1(x,y) (B-h_m)^{2 k+1} =F_B(x,y).\label{eq:2}
\end{eqnarray}

Let us apply the operator $\Delta_2$ to both sides of (\ref{eq:2}),
now harmonicity of the function
\begin{eqnarray}
u^m_{2 n+1}(x,y)=\frac{(-1)^{n}}{(2n+1)!}\Delta^{n}_2
u^m_1(x,y)\nonumber
\end{eqnarray}
implies
\begin{eqnarray}
\sum\limits_{k=0}^{n-1}\frac{(-1)^{k}}{(2k+1)!} \Delta_2^{k+1}
u^m_1(x,y) (B-h_m)^{2 k+1}=\Delta_2 F_B.\nonumber
\end{eqnarray}

When we apply the operator {\bf $\Delta_2$} to both sides of the
last relation, the function $u^m_{2 n-1}$ disappears from the left
part. We continue application of the operators $\Delta^k_2$ till
$k=n$. We obtain $n$ additional equalities from equality
(\ref{eq:2}). After we apply the operator  $\Delta^n_2$ to
(\ref{eq:2}) we have
$$
(B-h_m)\Delta_2^{n} u^m_1=(B-h_m)\frac{(2n+1)!}{(-1)^n} u^m_{2 n
+1}=\Delta_2^{n} F_B,\nonumber
$$
and it is the expression of the harmonic function $u^m_{2
n+1}(x,y)$.

When we apply $\Delta^{n-1}_2$ we have:
\begin{eqnarray}
(B-h_m) \Delta_2^{n-1} u_1^m+(B-h_m)^3\Delta_2^{n-1} u_3^m= & \nonumber \\
=(B-h_m)\frac{(2n-1)!}{(-1)^{n-1}}u_{2 n-1}^m-(B-h_m)^3\frac{1}{6}
\Delta_2^n u_1^m= & \nonumber \\
=(h_m-B)\frac{(2n-1)!}{(-1)^{n-1}}u_{2
n-1}^m-(B-h_m)^3\frac{(2n+1)!}{6 (-1)^n} u_{2n+1}^m= &
\Delta_2^{n-1} F_B.\nonumber
\end{eqnarray}

After we put the known expression of $u_{2n+1}^m$ into the last
relation  we get the biharmonic function $u^{m}_{2 n -1}$. We
continue our inverse movement over the additional equalities and
restore all functions  $u_k^m(x,y)$, beginning with
$u_{2n+1}^m(x,y)$ up to $u_1^m(x,y)$. We get the last function
directly from equality (\ref{eq:2}). When we restore all functions
$u_{2 k+1}^m(x,y)$, $k=1,...,n$, we have the last spline in the
layer $h \in [h_m,B]$.

The solution of the problem for a cylinder can be reduced to the
solution for the circular cylinder due to the additional conform
mapping.

Note that as at the previous section we can construct the solution
beginning from the lowest end as well as from the highest end. If
$u_l(x,y,h)$ and $u_h(x,y,h)$ are such solutions then $\alpha
u_l(x,y,h)+(1-\alpha) u_h(x,y,h)$ is also the solution for any
$\alpha \in(0,1)$.

\section{Example}

Let the domain $M$ be the half-ball bounded by the plane $XOY$ from
below and by the sphere $x^2+y^2+h^2=1$ from above. We construct the
simplest spline-interpolation solution with the given data on the
base $h=0$, on the level $h=1/2$ of the spherical surface and at the
vertex at the level $h=1$. We suppose that the data on the base is a
biharmonic function on the variables $x$ and $y$:
\begin{eqnarray}
u(x,y,0) = Re (\sum_{k=0}^N (a_k^0-i b_k^0)(x+i y)^k)+\nonumber\\
+ (x^2+y^2) Re (\sum_{k=0}^N (a_k^1-i b_k^1)(x+i y)^k).\nonumber
\end{eqnarray}

The data on the curve $\{x^2+y^2=\frac{3}{4}, h=\frac{1}{2}\}$ has
the form
\begin{eqnarray}
u(\frac{\sqrt3}{2} \cos \phi, \frac{\sqrt3}{2} \sin
\phi,\frac{1}{2}) = \frac{a_0}{2} + \sum_{k=0}^N (a_k \cos \phi +
b_k \sin \phi).\nonumber
\end{eqnarray}

The data at the vertex is $u(0,0,1)=u_0$.

We search for the solution for both layers $h \in [0,\frac{1}{2}]$
and $h \in [\frac{1}{2},1]$ in the form of the polynomial of the
second power over $h$.

In the layer $h \in [0,1/2]$ the spline has the form
\begin{eqnarray}
u(x,y,h)=u_0^1(x,y)+h u_1^1(x,y)+h^2 u_2^1(x,y),\nonumber
\end{eqnarray}
where
\begin{eqnarray}
u_0^1(x,y)=Re (\sum_{k=0}^N (a_k^0-i b_k^0)(x+i y)^k)+\nonumber\\
+ (x^2+y^2) Re (\sum_{k=0}^N (a_k^1-i b_k^1)(x+i y)^k),\nonumber
\end{eqnarray}
\begin{eqnarray}
u_1^1(x,y)=2 [\frac{a_0}{2} + Re (\sum_{k=1}^N (a_k-i b_k)(\frac{2}{\sqrt3})^k (x+i y)^k)-\nonumber\\
-Re (\sum_{k=0}^N (a_k^0-i b_k^0)(x+i y)^k)-\frac{a_0^1}{4}+\nonumber\\
+\frac{1}{4} Re (\sum_{k=1}^N (a_k^1-i b_k^1) (2 k-1)(x+i
y)^k)],\nonumber
\end{eqnarray}
\begin{eqnarray}
u_2^1(x,y)=-2 [a_0^1+Re (\sum_{k=1}^N (a_k^1-i b_k^1) (k+1)(x+i
y)^k)].\nonumber
\end{eqnarray}

In the layer $h \in [1/2,1]$ the spline has the form
\begin{eqnarray}
u(x,y,h)=u_0^2(x,y)+(h-\frac{1}{2}) u_1^2(x,y)+(h-\frac{1}{2})^2
u_2^2(x,y),\nonumber
\end{eqnarray}
where
\begin{eqnarray}
u_0^2(x,y)=\frac{a_0}{2}+ \mathrm{Re} (\sum_{k=1}^N (a_k-i b_k)(\frac{2}{\sqrt3})^k (x+i y)^k)-\nonumber\\
-\frac{3}{4} Re (\sum_{k=0}^N (a_k^1-i b_k^1)(x+i y)^k)+ (x^2+y^2)
\mathrm{Re} (\sum_{k=0}^N (a_k^1-i b_k^1)(x+i y)^k),\nonumber
\end{eqnarray}
\begin{eqnarray}
u_1^2(x,y)=2 u_0 + \frac{5}{2} a_0^1- a_0,\nonumber
\end{eqnarray}
\begin{eqnarray}
u_2^2(x,y)=-2 [a_0^1+ \mathrm{Re} (\sum_{k=1}^N (a_k^1-i b_k^1)
(k+1)(x+i y)^k)].\nonumber
\end{eqnarray}

\section{Approximation estimate of the continuous solution}

Assume that $\tilde{u}$ is the difference between the exact solution
of Dirichlet problem and the spline-interpolation solution presented
in the previous sections. This function vanishes at the points of
$C_j$, at the points of the ends or at the top points. The function
$\tilde{u}(x,y,h)$ has continuous derivatives $\tilde{u}'_x(x,y,h)$
and $\tilde{u}'_y(x,y,h)$ at any intersection of the plane $h=const$
with $M$. So we have the estimate
$$
|\tilde{u}(x,y,\tilde{h})| \leq \hat{u}(\tilde{h}) + l
\max\limits_{(x,y,h)\in M}
\sqrt{\tilde{u}'^2_x(x,y,h)+\tilde{u}'^2_y(x,y,h)},
$$
where $\hat{u}(\tilde{h})$ is the maximum of the module of the
boundary difference between the classic solution and the constructed
approximate solution on the level $h=\tilde{h}$, $l$ is the distance
from the point $(x,y)$ to the boundary of the intersection of $M$
with the plane $h=\tilde{h}$. Evidently, $\hat{u}(\tilde{h})=0$ for
$\tilde{h}=h_j,$ $j=1,2,...,m$. The values of $\hat{u}(\tilde{h})$
tend to zero when we increase the number of $m$ and decrease the
value $\max\limits_j|h_{j+1}-h_j|$ if the corresponding classic
Dirichlet problem has smooth boundary values.

The estimate shows that the continuous solution is rather accurate
in the neighbourhood of the boundary $S$ of the body $M$.

\section{Construction of the smoothing solution}

Here we construct the continuous solution which minimizes the
difference between the derivatives with respect to $h$ of the
adjacent splines at the common disk $E_j$. We consider a body of
revolution without the ends or with one end. We suppose that the
function $\Phi''(h)$ is continuous and the boundary data of the
corresponding classic Dirichlet problem $f(\theta,h)$ has the form
$\sum\limits_{k=0}^l C_k(t)\cos k\theta +D_k\sin k\theta$ where the
functions $C_k(t)$ and $D_k(t)$ are at least $3$-differentiable,
$h\in[A,B]$.

The spline for the layer $h\in[h_j,h_{j+1}]$ has the following form:
$$
u^{j+1}(x,y,h)=\sum\limits_{k=0}^{2N+1} u_{k}^{j+1}(x,y) h^k,
$$
where
$$
u_{2p}^{j+1}(x,y)=(-1)^p \Delta^p u_0^{j+1}(x,y)/(2p)!,
$$
$$
u_{2p+1}^{j+1}(x,y)=(-1)^p \Delta^p u_1^{j+1}(x,y)/(2p+1)!,\
p=0,...,N.
$$

We choose the $(N+1)$-harmonic coefficient $u_0^{j+1}(x,y)$ so that
$u^{j+1}(x,y,h_j)=u^j(x,y,h_j)$, hence all even coefficients
$u_{2p}^{j+1}(x,y)$, $p=0,...,N$, are known. We choose the
$(N+1)$-harmonic coefficient $u_1^{j+1}(x,y)$ so that
$$
u^{j+1}(\Phi(h_{j+1})\cos\theta,\Phi(h_{j+1})\sin\theta,h_{j+1})=f(\theta,h_{j+1})
$$
and the norm of the difference $u^{j+1}(x,y,h_j)-u^{j}(x,y,h_j)$ in
the space $L_2$ is close to minimal.

Let $h_j=0$, $h_{j-1}=-t$, $h_{j+1}=t$, $\Phi(h_j)=1$,
$\Phi(h_{j+1})=1+ t b(t)$, where $b(t)$ is bounded when $t$ tends to
zero. Gluing of the neighbor splines at the unit disk yields the
equation $u_{2p}^{j+1}(x,y)$ $= u_{2p}^{j}(x,y)$.

After we consider the boundary values  $u^{j+1}(x,y,t)$ and
$u^{j}(x,y,-t)$ and apply Lagrange formula we obtain in the same way
as for the liner splines the representation
\begin{eqnarray}
\sum\limits_{p=0}^N [u_{2p+1}^{j+1}(\cos\theta,\sin\theta) -
u_{2p+1}^{j}(\cos\theta,\sin\theta)]t^{2p+1} = t^2
\phi(\theta,t),\label{eq:3}
\end{eqnarray}
where $\phi(\theta,t)$ is bounded when $t$ tends to zero.

Consider the function
$$
u(x,y,h)=\sum\limits_{p=0}^N [u_{2p+1}^{j+1}(\cos\theta,\sin\theta)
- u_{2p+1}^{j}(\cos\theta,\sin\theta)]h^{2p+1}=
$$
$$
=\sum\limits_{p=0}^N u_{2p+1}(x,y) h^{2p+1}
$$
in the intersection of $M$ with the layer $h\in[0,t]$.

Let us introduce the polar coordinates and take the coefficient
$u_1(x,y)=u_1(r,\theta)$ in the  form of the following expansions
$$
u_1(r,\theta)=\sum\limits_{p=0}^{N}\sum\limits_{k=0}^l(\alpha_k^{2p}
\cos k\theta+\beta_k^{2p}\sin k\theta) r^{k+2p}.
$$

The last relation provides us with the expansions of the
coefficients $u_{2p+1}(r,\theta)$, $p=1,...,N$. Due to
representation (\ref{eq:3}) we can take the boundary data on the
level $h=t$ in the form
$$
u(\Phi(t),\theta)=t^2\sum_{k=0}^l(\lambda_k\cos k\theta+\mu_k\sin
k\theta) + t^3\psi(t,\theta).
$$

Therefore we have the relation
$$
t[\sum\limits_{k=0}^l(\alpha_k^0 \cos k\theta+\beta_k^0 \sin
k\theta)(1+t b(t))^k+\sum\limits_{k=0}^l(\alpha_k^2 \cos
k\theta+\beta_k^2 \sin k\theta)(1+t b(t))^{k+2}+...
$$
$$
+\sum\limits_{k=0}^l(\alpha_k^{2N} \cos k\theta+\beta_k^{2N} \sin
k\theta)(1+t
b(t))^{k+2N}]-\frac{t^3}{3!}[\sum\limits_{k=0}^l4(k+1)(\alpha_k^2
\cos k\theta+
$$
$$
+\beta_k^2 \sin k\theta)(1+t b(t))^k+...+\sum\limits_{k=0}^l4 N
(k+N)(\alpha_k^{2N} \cos k\theta+\beta_k^{2N} \sin k\theta)(1+t
b(t))^{k+2N-2}]+
$$
$$
+...+(-1)^N\frac{t^{2N+1}}{(2N+1)!}\sum\limits_{k=0}^l 4^N
N!\frac{(k+N)!}{k!}(\alpha_k^{2N} \cos k\theta+\beta_k^{2N} \sin
k\theta)(1+t b(t))^k=
$$
$$
=t^2\sum_{k=0}^l(\lambda_k\cos k\theta+\mu_k\sin k\theta) +
t^3\psi(t,\theta).
$$

The last relation yields the following representation of the
coefficients $\alpha_k^0,\beta_k^0$ through the coefficients
$\alpha_k^{2j},\beta_k^{2j}$, $k=0,..,l$, $j=2,...,N$,
\begin{eqnarray}
\alpha_k^0=t\lambda_k -\alpha_k^2-\alpha_k^4-...-\alpha_k^{2N}+t
\sum\limits_{j=1}^N c_k^j \alpha_k^{2j}+t^2 \phi_k(t),\label{eq:4}
\end{eqnarray}
\begin{eqnarray}
\beta_k^0=t\mu_k
-\beta_k^2-\beta_k^4-...-\beta_k^{2N}+\sum\limits_{j=1}^N d_k^j
\beta_k^{2j}+t^2 \psi_k(t),\label{eq:5}
\end{eqnarray}
where $\phi_k(t)$ and
$\psi_k(t)$ are bounded when $t$ tends to zero.

Now we must choose the coefficients $\alpha_k^{2j},\beta_k^{2j}$ so
that the value $\|u_1\|_{L_2(E)}$ becomes rather small. Here
\begin{eqnarray}
\|u_1\|^2_{L_2(E)}=2\pi\int\limits_0^1[\alpha_0^0+\alpha_0^2
r^2+...+\alpha_0^{2N} r^{2N}]^2 r dr +  \pi
\sum\limits_{k+1}^{l}[\int\limits_0^1(\alpha_k^0+\nonumber
\end{eqnarray}
\begin{eqnarray}
+\alpha_k^2 r^{2+k}+...+\alpha_k^{2N} r^{2N+k})^2 r
dr+\int\limits_0^1(\beta_k^0+\beta_k^2 r^{2+k}+...+\beta_k^{2N}
r^{2N+k})^2 r dr].\label{eq:6}
\end{eqnarray}

{\bf Lemma 1.} The minimum of the quadratic form
$$
\int\limits_0^1[a_0+a_1 r^2+...+a_N r^{2N}]^2 r dr
$$
with the condition $\sum\limits_{k=0}^N a_k=1$ is equal to
$\frac{1}{2(N+1)^2}$.

{\bf Proof}. We introduce the variable $t=r^2$ and represent the
expression $a_0+a_1 t+...+a_N t^N$ as linear combination of
polynomials $\hat{P}_k(t)=P_k(2t-1)$, $t\in[0,2\pi]$, where $P_k(t)$
are Legendre polynomials:
$$
a_0+a_1 t+...+a_N t^N=\sum\limits_{k=0}^N b_k \hat{P}_k(t).
$$
This representation is unique and gives us the linear relations
between the coefficients $a_k$ and $b_k$, $k=0,...,N$. Now the given
condition takes the form
$$
\sum\limits_{k=0}^N a_k=\sum\limits_{k=0}^N
b_k\hat{P}_k(1)=\sum\limits_{k=0}^N b_k=1.
$$
and the given quadratic form is equal to
$$
\frac{1}{2}\int\limits_0^1(\sum\limits_{k=0}^N b_k \hat{P}_k(t))^2
dt =\frac{1}{2}\sum\limits_{k=0}^N \frac{b_k^2}{2k+1}.
$$

We apply Lagrange method and find the values
$\tilde{b}_k=\frac{2k+1}{(N+1)^2}$ which provide minimum of the
corresponding quadratic form with the given condition. So the
minimum of quadratic form equals to
$$
\frac{1}{2}\sum\limits_{k=0}^N
\frac{\tilde{b}_k^2}{2k+1}=\frac{1}{2(N+1)^2}.\square
$$

We denote by $\tilde{a}_k^N$, $k=0,...,N$ the values of the
coefficients which provide minimum of the quadratic form given in
the statement of Lemma 1.

Now we put $\alpha_0^{2p}= t\lambda_0 \tilde{a}_p^N,$ $p=1,...,N$.
So the first summand from (\ref{eq:6}) according to (\ref{eq:4})
with $k=0$ has the following representation
$$
2\pi\int\limits_0^1[\alpha_0^0+\alpha_0^2 r^2+...+\alpha_0^{2N}
r^{2N}]^2 r dr =
2\pi\int\limits_0^1\{t\lambda_0[1-(\tilde{a}_1^N+...+\tilde{a}_N^N)+
$$
$$
+\tilde{a}_1^N r^2+...+\tilde{a}_N^N r^{2N}]+t^2 R_0(r,t,N)\}^2 r
dr=\frac{\pi t^2 \lambda_0^2}{(N+1)^2}[1+t \tilde{R}_0(t,N)].
$$

We need to find  minima of some other quadratic forms but we have to
introduce some polynomial bases different from Legendre polynomials
in order to find the exact value of these minima. So we give here
only the estimates of them.

{\bf Lemma 2.} The minimum of the quadratic form
$$
\int\limits_0^1[a_0+r^k(a_1 r^2+...+a_N r^{2N})]^2 r dr
$$
with the condition $\sum\limits_{k=0}^N a_k=1$ is less than
$\frac{1}{2 N^2}$.

{\bf Proof.} We have
$$
\int\limits_0^1[a_0+r^k(a_1 r^2+...+a_N r^{2N})]^2 r
dr=\frac{a_0}{2}+2 a_0 \int\limits_0^1 r^{k+1}(a_1 r^2+...+a_N
r^{2N}) dr +
$$
$$
+\int\limits_0^1 r^{2k+4}(a_1+...+a_N r^{2N-2})^2 r dr.
$$

Let us put $a_0=0$. Now
$$
\int\limits_0^1[a_0+r^k(a_1 r^2+...+a_N r^{2N})]^2 r
dr=\int\limits_0^1 r^{2k+4}(a_1+...+a_N r^{2N-2})^2 r dr\leq
$$
$$
\leq \int\limits_0^1 (a_1+...+a_N r^{2N-2})^2 r dr.
$$

The minimum of the  quadratic form $\int\limits_0^1 (a_1+...+a_N
r^{2N-2})^2 r dr$ according to Lemma 1 is equal to $\frac{1}{2
N^2}$. This quadratic form achieves the minimal value when
$a_k=\tilde{a}_{k-1}^{N-1}$, $k=1,...,N$. Lemma 2 is
proved.$\square$

Now we put
$$
\alpha_k^{2p}=\lambda_k t\tilde{a}_{p-1}^{N-1},
$$
$$
\beta_k^{2p}=\mu_k t\tilde{a}_{p-1}^{N-1},\   \  p=1,...,N,
k=1,...,l.
$$
and obtain according to (\ref{eq:4}) and (\ref{eq:5}) the estimates
of the following summands from (\ref{eq:6}):
$$
\pi \int\limits_0^1(\alpha_k^0+\alpha_k^2 r^{2+k}+...+\alpha_k^{2N}
r^{2N+k})^2 r dr \leq \frac{\pi \lambda_k^2 t^2}{2
N^2}(1+\tilde{R}_k(t,N)),
$$
$$
\pi \int\limits_0^1(\beta_k^0+\beta_k^2 r^{2+k}+...+\beta_k^{2N}
r^{2N+k})^2 r dr \leq \frac{\pi \mu_k^2 t^2}{2
N^2}(1+\tilde{Q}_k(t,N)).
$$

Finally
$$
\|u_1\|^2_{L_2(E)}\leq \frac{A t^2}{N^2} +t^3\tilde{\Phi}(t,N),
$$
where $A$ is constant and $\tilde{\Phi}(t,N)$ is bounded for the
fixed $N$ when $t$ tends to zero and therefore
\begin{eqnarray}
\|u^{j+1}_h-u^j_h\|^2_{L_2(E_j)}\leq \frac{C t^2}{N^2}
+t^3\tilde{\Phi}(t,N).\label{eq:7}
\end{eqnarray}

\section{Approximation estimate for the smoothing solution}

In this section we present integral approximation estimate of the
constructed solution. We suppose that the distance $|h_{j+1}-h_j|$
is unique for all $j=1,...,m,$ and equals to $t=(B-A)/(m+1)$.

Assume that $\hat{u}$ is the difference between the exact solution
of Dirichlet problem and the smoothing spline-interpolation
solution. Let us first consider the equality which holds true due to
Gauss-Ostrogradsky theorem
$$
\int\int\limits_{S} \hat{u} (\hat{u}_h dx dy+ \hat{u}_x dy dh+
\hat{u}_y dh dx)+\sum\limits_{j=1}^m  \int\int\limits_{E_j} \hat{u}
(u^j_h(x,y,h_j)-u^{j+1}_h(x,y,h_j)) dx dy =
$$
$$
 =\int\int\limits_{M}\int |\mathrm{grad} \hat{u}(x, y,
h)|^2 dx dy dh,
$$
where $u^j(x,y,h)$ is the spline at the layer $h\in[h_{j-1},h_j]$
and $E_j$ is the disk which is the intersection of $M$ with the
plane $h=h_j$.

So
$$
\int\int\limits_{M}\int |\mathrm{grad} \hat{u}(x, y, h)|^2 dx dy dh
\leq \max_{S} |\hat{u}| \int\int\limits_{S}  |\mathrm{grad} \hat{u}
|dS +
$$
$$
+\max_{M} |\hat{u}| \sum\limits_{j=1}^m  \int\int\limits_{E_j}
|(u^j_h(x,y,h_j)-u^{j+1}_h(x,y,h_j))| dx dy.
$$

The right-hand side of the last inequality  can be made arbitrary
small due to the approximation of the boundary data and the
smoothing of the derivatives with respect to $h$. Really
$$
\sum\limits_{j=1}^m  \int\int\limits_{E_j}
|(u^j_h(x,y,h_j)-u^{j+1}_h(x,y,h_j))| dx dy \leq \frac{D m(B-A)}{N
(m+1)} +
$$
$$
+ m \frac{(B-A)^2}{(m+1)^2}\tilde{\Psi}(t,N),
$$
where $D$ is constant and $\tilde{\Psi}(t,N)$ is bounded when $t$
tends to zero, due to (\ref{eq:7}) and to H\"older inequality. Let
$\max_{M} |\hat{u}|>0$. For arbitrary small $\varepsilon >0$ we can
choose $N$ so large that the first summand in the right side of the
previous inequality is less than $\varepsilon/(3 \max_{M}
|\hat{u}|)$. Now with the chosen number $N$ we can choose the number
$m$ so large that the second summand in the right side of the
previous inequality is less than $\varepsilon/(3 \max_{M}
|\hat{u}|)$.

Obviously  the value of $\max_{S} |\hat{u}| [\int\int\limits_{S}
|\mathrm{grad} \hat{u} |dS$ can be made less than $\varepsilon/3$
for sufficiently large number $m$.

Therefore $\int\int\limits_{M}\int |\mathrm{grad} \hat{u}(x, y,
h)|^2 dx dy dh < \varepsilon$ for large values of $N$ and $m$. Hence
the H\"older inequality implies
$$
\int\int\limits_{M}\int |\mathrm{grad} \hat{u}(x, y, h)|dx dy dh
\leq \sqrt{V(M) \varepsilon},
$$
where $V(M)$ is the volume of $M$.

Now we prove  the following estimate:
$$
\int\int\limits_{P_{\tilde{h}}}|\hat{u}(x,y,\tilde{h})| dx dy \leq Q
\sqrt{\varepsilon}
$$
for any $\tilde{h}\in [A,B]$, where $E_{\tilde{h}}$  is the section
of $M$ by the plane $h=\tilde{h}$.

We prove this estimate at first for the case when  $M$ is a convex
body of revolution.

Let us fix $\tilde{h}$ and choose a point $s^*\in U$, where
$U=\bigcup_{j=1}^n C_j\bigcup C_A \bigcup C_B$;  $C_j$, $j=1,...,n$,
are the curves with the given boundary data on $S$ and $C_A$, $C_B$
are the corresponding edges (or points). The cone $CH(P_h,s^*)$
which is the convex hull of $P_h \bigcup \{s^*\}$ is a subset of $M$
since $M$ is convex.  Let us consider the set $\{\gamma_{\tilde{h}}
(x,y,s^*)\}$ of the straight lines, connecting the points $(x,y)$ of
$P_{\tilde{h}}$ with the point $s^*$ and parametrized linearly by
$h$:
$$
\gamma_{\tilde{h}} (x,y,s^*)=(x(h),y(h),h), \   \  h \in
[\tilde{h},h^*],
$$
where $h^*$ is the $h$-coordinate of $s^*$.   One can find the
characteristic
$$
\tilde{K}(\tilde{h},s^*)=\max_{(x,y)\in P_{\tilde{h}}}|\frac{d
\gamma_{\tilde{h}} (x,y,s^*)}{d h}|.
$$

Now we find the characteristic $K(\tilde{h})=\min_{s^*\in U}
\tilde{K}(\tilde{h},s^*)$ and the point $s^*(\tilde{h})$ which
realises this minimum, the $h$-coordinate of the point
$s^*(\tilde{h})$ being $h^{**}$.

The values of the function $\hat{u}$ vanish on the set $U$, so for
any point $p\in P_{\tilde{h}}$ we have
$$
u(p)=\int\limits_{\tilde{h}}^{h^{**}} \frac{d \hat{u}(x(h), y(h),
h)}{dh} dh=\int\limits_{\tilde{h}}^{h^{**}} (\mathrm{grad}
\hat{u}(x, y, h), \frac{d\gamma_{\tilde{h}}}{dh}) dh.
$$
Hence the estimate of $\frac{d\gamma_{\tilde{h}}}{d h}$ implies
$$
|\hat{u}(p)|\leq K(\tilde{h}) |\int\limits_{\tilde{h}}^{h^{**}}
\|\mathrm{grad} \hat{u}(x, y, h)\| dh|.
$$
Thus
$$
\int\int\limits_{P_{\tilde{h}}} |\hat{u}(x,y,\tilde{h})| dx
dy=\int\int\limits_{CH(P_{\tilde{h}},s^*(\tilde{h}))}\int
|(\mathrm{grad} \hat{u}(x, y, h), \frac{d\gamma}{dh})| dx dy d h
\leq
$$
$$
\leq K(\tilde{h}) \int\int\limits_{M}\int |\mathrm{grad}
\hat{u}(x,y,h)|dx dy dh \leq K(\tilde{h}) \sqrt{V(M) \varepsilon} =
Q \sqrt{\varepsilon}.
$$
Here $Q=\sqrt{V(M)}\max\limits_{\tilde{h} \in [A,
B]}\{K(\tilde{h})\}$.

For another types of bodies we can obtain similar estimate if we
divide any section $P_{\tilde{h}}$ into finite number of domains,
construct the cone with the vertex from $U$ which lies in $M$ for
every domain and take the integral of $|\hat{u}|$ over
$P_{\tilde{h}}$ as the sum of the integrals over the introduced
domains.

\section{On the similar problems}

As it was noted above the external problem for the body of
revolution can be reduced to the internal one when the unbounded
domain in which we solve the problem contains any ball in its
exterior. We make the inversion with respect to the corresponding
sphere and obtain the internal problem for the bounded domain. We
construct the spline-interpolation solution of this internal problem
and then make the inversion once more.

The most simple case of the external problem is the case of the
problem for the  unit ball $B_1(0)$ exterior. The points of the unit
sphere are invariant under the inversion. So the boundary conditions
remain the same.

We also can apply the spline-interpolation solution presented in
this paper to the Dirichlet value problem for a cylinder or a cone
with the section which is the known conformal map of the unit disk.
Solution of 2D Dirichlet problem for the corresponding cuts can be
reduced to solution of this problem in the unit disk.


\end{document}